\documentclass[11pt]{article}

\usepackage{epsf,latexsym,amssymb}
\usepackage{times}
\usepackage{graphicx}
\usepackage{amsmath}  

\newcommand{\FIGDIR}{iffl_jan2016_figdir}
\newcommand{\pic}[2]{\includegraphics[scale=#1]{\FIGDIR/#2}}
\newcommand{\picc}[2]{\begin{center}\pic{#1}{#2}\end{center}}

\newcommand{\comment}[1]{}
\newcommand{\edo}{\end{document}}

\newcommand{\twoif}[4]{
\left\{ \begin{array}{ll}#1&#2\\#3&#4\end{array}\right.
}

\newtheorem{theorem}{Theorem}
\newtheorem{itlemma}{Lemma}[section] 
\newtheorem{itproposition}[itlemma]{Proposition}
\newtheorem{itcorollary}[itlemma]{Corollary}
\newtheorem{itremark}[itlemma]{Remark}
\newtheorem{itdefinition}[itlemma]{Definition}
\newtheorem{itexample}[itlemma]{Example}

\newenvironment{lemma}{\begin{itlemma}\rm}{\end{itlemma}} 
\newenvironment{remark}{\begin{itremark}\rm}{\end{itremark}} 
\newenvironment{corollary}{\begin{itcorollary}\rm}{\end{itcorollary}}
\newenvironment{proposition}{\begin{itproposition}\rm}{\end{itproposition}}
\newenvironment{definition}{\begin{itdefinition}\rm}{\end{itdefinition}}
\newenvironment{example}{\begin{itexample}\rm}{\end{itexample}}

\newcommand{\be}[1]{\begin{equation}\label{#1}}
\newcommand{\ee}{\end{equation}}
\newcommand{\bl}[1]{\begin{lemma}\label{#1}}
\newcommand{\ble}[1]{\begin{lemmaex}\label{#1}}
\newcommand{\br}[1]{\begin{remark}\label{#1}}
\newcommand{\bt}[1]{\begin{theorem}\label{#1}}
\newcommand{\bd}[1]{\begin{definition}\label{#1}}
\newcommand{\bp}[1]{\begin{proposition}\label{#1}}
\newcommand{\bc}[1]{\begin{corollary}\label{#1}}
\newcommand{\bex}[1]{\begin{example}\label{#1}}
\newcommand{\ec}{\mybox\end{corollary}}
\newcommand{\eex}{\mybox\end{example}}
\newcommand{\eem}{\mybox\end{example}}
\newcommand{\el}{\mybox\end{lemma}}
\newcommand{\er}{\mybox\end{remark}}
\newcommand{\et}{\qed\end{theorem}}
\newcommand{\ed}{\mybox\end{definition}}
\newcommand{\ep}{\mybox\end{proposition}}
\newcommand{\epr}{\end{proof}}
\newcommand{\bpr}{\begin{proof}}

\newcommand{\ecs}{\end{corollary}}
\newcommand{\eexs}{\end{example}}
\newcommand{\els}{\end{lemma}}
\newcommand{\ers}{\end{remark}}
\newcommand{\ets}{\end{theorem}}
\newcommand{\eds}{\end{definition}}
\newcommand{\eps}{\end{proposition}}
\newcommand{\halmos}{\rule{1ex}{1.4ex}}
\newcommand{\qed}{\hfill \halmos} 
\newcommand{\mybox}{\hfill $\Box$} 
\newcommand{\beq}{\begin{eqnarray}}
\newcommand{\eeq}{\end{eqnarray}}
\newcommand{\beqn}{\begin{eqnarray*}}
\newcommand{\eeqn}{\end{eqnarray*}}
\newcommand{\bi}{\begin{itemize}}
\newcommand{\ei}{\end{itemize}}
\newcommand{\ben}{\begin{enumerate}}
\newcommand{\een}{\end{enumerate}}

\newcommand{\bes}[1]{\begin{subequations}\label{#1}\begin{eqnarray}}
\newcommand{\ees}{\end{eqnarray}\end{subequations}}

\newenvironment{proof}{\noindent {\em Proof}.\ }{\hspace*{\fill}$\halmos$\medskip}

\newcommand{\mypmatrix}[1]{\left(\begin{array}{cccccccccccc}#1\end{array}\right)}

\newcommand{\sysa}{(\ref{sys:production.expo}a)}
\newcommand{\sysb}{(\ref{sys:production.expo}b)}
\newcommand{\sysc}{(\ref{sys:production.expo}c)}
\newcommand{\sysab}{(\ref{sys:production.expo}ab)}
\newcommand{\sysabc}{(\ref{sys:production.expo}abc)}

\newcommand{\sysabpc}{(\ref{sys:production.expo}ab'c)}
\newcommand{\sysabp}{(\ref{sys:production.expo}ab')}

\newcommand{\sysfeedbackab}{(\ref{eq:full_feedback_system}ab)}
\newcommand{\sysfeedbackabc}{(\ref{eq:full_feedback_system}abc)}

\newcommand{\xstar}{x^*}
\newcommand{\ystar}{y^*}
\newcommand{\tstar}{t^*}

\newcommand{\muinf}{\underline{\mu }}
\newcommand{\musup}{\overline{\mu }}
\newcommand{\pinf}{\underline{p}}
\newcommand{\psup}{\overline{p}}

\newcommand{\ybar}{\bar{y}}
\newcommand{\pbar}{\bar{p}}
\newcommand{\vbar}{\bar{v}}

\title{A remark on incoherent feedforward circuits \\
  as change detectors and feedback controllers}
\author{Eduardo D. Sontag\\
Rutgers Uninversity}

\begin{document}
\maketitle

\begin{abstract}

This note analyzes incoherent feedforward loops in signal processing and
control.  It studies the response properties of IFFL's to exponentially
growing inputs, both for a standard version of the IFFL and for a variation in
which the output variable has a positive self-feedback term.  It also
considers a negative feedback configuration, using such a device as a
controller.  It uncovers a somewhat surprising phenomenon in which
stabilization is only possible in disconnected regions of parameter space, as
the controlled system's growth rate is varied.

\end{abstract}

\section{Introduction}

This note derives several theoretical results regarding the use of incoherent
feedforward loops (IFFL's) in signal processing and control.
We will study the system:
\bes{sys:production.expo}
\dot x &=& -ax + bu\\
\dot y &=& c\frac{u}{x} - \delta y\\
\dot u &=& (\lambda -\kappa y)u
\ees
as well as a modified system in which there is also an autocatalytic term in 
\sysb:
\be{eqn:yfeedback}
\tag{\ref{sys:production.expo}b'}
\dot y \;=\; c\frac{u}{x} \,-\, \delta y \,+\, \frac{V y^n}{K^n + y^n}
\ee
which represents a positive feedback of the $y$ variable on itself.
The constants $a,b,c,\delta ,\kappa ,V,K$ are positive
(but $\lambda $ is allowed to be negative), dot indicates $d/dt$, 
$n$ is typically an integer$>1$ that represents molecular cooperativity,
and the scalar functions of time $x=x(t)$, $y=y(t)$, and $u=u(t)$
take positive values.  (It is easy to verify that, for any positive initial
conditions, solutions remain positive for all times.)  Of course, setting
$V=0$ allows seeing \eqref{eqn:yfeedback} as a special case of \sysb,
but it is more interesting to treat the non-autocatalytic case by itself.

We will separately study the first two equations \sysab{} (or
\sysabp{} when there is an autocatalytic term),
viewing $u=u(t)$ as an external input to the IFFL described by
\sysab{} (or \sysabp),
and viewing $y=y(t)$ as an output or response of the system.
Later, we ``close the loop'' by letting $u$ be described by
(\ref{sys:production.expo}c), thinking of it as a variable that is
controlled by $y$ through a negative feedback with gain $\kappa $, and which,
conversely, feeds back into the IFFL through the $x$ variable.
In that context, we study the full system \sysabc{} (or \sysabpc). 

The motivation for this work is the potential role that these motifs might
play in immunology \cite{biorxiv_change_detection_immune_2015}.
In that context, one might view the $x$ variable as representing the level of
activity of a regulatory inhibitory component (such as a population of
T$_{\mbox{reg}}$ cells at a particular infection site or in a certain tumor
microenvironment), $y$ as the level of activity of an immune response component
(such as cytotoxic T cells), and $u$ as a population of pathogens or the
volume of a tumor, which might grow exponentially (if $\lambda >0$) in the absence
of immune action, but which is killed at a rate proportional to the immune
response.  The feedback into $x$ and $y$ represents the activation of both the
response and of the regulatory mechanism in response to the infection or tumor.
As remarked in \cite{biorxiv_change_detection_immune_2015}, a very interesting
feature of the IFFL controller is its capability of detecting change as well
as the fact that the level of activity is proportional to the rate of growth
of the input, which may account for tolerance of slow-growing infections and
cancers as well as Weber-like logarithmic sensing and ``fold change
detection'' of inputs.

In an immunological context, autocatalytic feedback might be implemented by
a cytokine-mediated recruiting of additional immune components, or by
autocrine stimulation.  This results in an excitable system, which allows $y$
to ``lock'' into a high state of activity given a sufficiently rapid rate of
change in its input.
Changing the growth rate $\lambda $ of the pathogen or tumor, while fixing
all other parameters, results in elimination of $u$ for small growth rates
$\lambda $, and in proliferation as $\lambda $ increases.  This is, of course, obvious.
However, and very surprisingly, it may happen in this model that further
increase of the growth rate $\lambda $, that is, when presented with a more aggressive
pathogen or tumor, leads to the eventual elimination of the pathogen or tumor.
This might be intuitively interpreted as a higher growth rate triggering
locking of the immune response at a higher value.  
An even larger increase in $\lambda $ leads again to proliferation.
In other words, the pattern ``elimination, proliferation, elimination,
proliferation'' can be obtained simply by gradually increasing $\lambda $.

\br{rem:rescaling}
In the system \sysabc, and in particular in the system
\sysab, one may assume without loss of generality that $a=b=c=1$.
This is because we may eliminate these parameters by rescaling variables.
Indeed, substituting
\be{rescaling:production}
x = \frac{b}{a}\xstar\,,\;
y = \frac{c}{b}\ystar\,,\;
t = \frac{1}{a}\tstar\,,\;
\delta ^* = \frac{\delta }{a}\,,\;
\lambda ^* = \frac{\lambda }{a}\,,\;
\kappa ^* = \frac{c\kappa }{ab}\,,
\ee
into system \sysabc{}, one obtains:
\bes{sys:production.expo.fixabc1}
\frac{d\xstar}{d\tstar} &=& -\xstar + u\\
\frac{d\ystar}{d\tstar} &=& \frac{u}{\xstar} - \delta ^*\ystar\\
\frac{du}{d\tstar}  &=& (\lambda -\kappa \ystar)u
\ees
\er

\section{IFFL's responses to various classes of inputs}

Let us consider the system \sysab, a differentiable function
$u=u(t)$ viewed as an external input or forcing function, and any (positive)
solution $(x(t),y(t))$ corresponding to this input.
We are interested first in understanding how the growth rate of the input
affects the asymptotic values of the output variable $y$.

We denote the derivative of  $\ln u(t)$ with respect to $t$ as follows:
\[
v(t)\;:=\; \frac{\dot u(t)}{u(t)}
\]
and its limsup and liminf as $t\rightarrow \infty $
\[
\muinf = \liminf_{t\rightarrow \infty } v(t)\,,\quad
\musup = \limsup_{t\rightarrow \infty } v(t)\,.
\]
We assume that $v$ is bounded, and thus both of these numbers are finite.
We also introduce the following function:
\[
p(t)\;:=\; \frac{u(t)}{x(t)}\,.
\]
Since
\[
\dot p = \dot u/x - u\dot x/x^2 = (u/x)[\dot u/u - \dot x/x] = 
(u/x)[\dot u/u - (-ax+bu)/x] = (u/x)[\dot u/u +a - bu/x] \,,
\]
we have that $p$ satisfies the following ODE with input $v$: 
\be{eq:pv}
\dot p = p (a+v-bp) \,.
\ee

\bl{lemma:plims}
Let $u$ be a differentiable input to system \sysab{} with $a=b=c=1$.
With the above notations,
\be{eq:plims}
\max\{0,1+\muinf\} \;\leq \; \liminf_{t\rightarrow \infty } p(t) 
\;\leq \; \limsup_{t\rightarrow \infty } p(t) \;\leq \; \max\{0,1+\musup\}
\ee
\els

\bpr
Since $a=b=c=1$,
\[
\dot p = p (1+v-p) \,.
\]
To prove the upper bound, we consider two cases,
$1+\musup<0$ and $1+\musup\geq 0$.
In the first case, 
let $\varepsilon :=-(1+\musup)>0$; the definition of $\musup$ gives that,
for some $T\geq 0$,
$1+v(t)< -\varepsilon /2$ for all $t\geq T$.
It follows that $\dot p\leq p(-\varepsilon /2-p)$ for all $t\geq T$. 
Thus, $\dot p<0$ whenever $p>0$, from which it follows that
$\limsup_{t\rightarrow \infty } p(t) = \lim_{t\rightarrow \infty } p(t) = 0$.
Suppose now that $1+\musup\geq 0$.
Pick any $\varepsilon >0$ and a $T=T(\varepsilon )\geq 0$ such that $v(t)\leq \musup+\varepsilon $ for all $t\geq T$.
For such $t$, $\dot p = p(1+v-p) \leq  p(1+\musup+\varepsilon -p)$.
This implies that $\dot p<0$ whenever $p(t)>1+\musup+\varepsilon $, which implies
that $\limsup_{t\rightarrow \infty }p(t)\leq 1+\musup+\varepsilon $.
Letting $\varepsilon \rightarrow 0$, we conclude that $\limsup_{t\rightarrow \infty }p(t)\leq 1+\musup$.

We next prove the lower bound.
Pick any $\varepsilon >0$ and a $T=T(\varepsilon )\geq 0$ such that $v(t)\geq \muinf-\varepsilon $ for all $t\geq T$.
Thus $\dot p = p(1+v-p) \geq  p(1+\muinf-\varepsilon -p)$ for all $t\geq T$.
This implies that $\dot p>0$ whenever $p(t)<1+\muinf-\varepsilon $
(recall that $p(t)>0$ for all $t$, since by assumption $u(t)>0$ and $x(t)>0$
for all $t$).
Therefore $\liminf_{t\rightarrow \infty }p(t)\geq 1+\muinf-\varepsilon $, and letting $\varepsilon \rightarrow 0$ we have
$\liminf_{t\rightarrow \infty }p(t)\geq 1+\muinf$.
Since $p(t)\geq 0$ for all $t$, we also have
$\liminf_{t\rightarrow \infty }p(t)\geq \max\{0,1+\muinf\}$.
This completes the proof.
\epr

In particular, if $v(t)\rightarrow \mu $ as $t\rightarrow \infty $ then $\muinf=\musup=\mu $, so we have as
follows. 

\bc{cor:plim}
If $v(t)\rightarrow \mu $ as $t\rightarrow \infty $ then 
$\displaystyle\lim_{t\rightarrow \infty } p(t) = \max\{0,1+\mu \}$.
\ec

For the original system \sysab, we have as follows.

\bp{prop:sysablimits}
Consider a solution of \sysab, with a differentiable $u(t)>0$ as input and
$x(t)>0$, $y(t)>0$.  Assuming that $v=\dot u/u$ is bounded, we have:
\be{eq:ylims}
\frac{c}{b\delta } \max\left\{0,a+\muinf\right\} \;\leq \; \liminf_{t\rightarrow \infty } y(t) 
\;\leq \; \limsup_{t\rightarrow \infty } y(t) \;\leq \; \frac{c}{b\delta }\max\left\{0,a+\musup\right\}
\ee
\eps
\bpr
We first assume that $a$$=$$b$$=$$c$$=$$1$.
Let $\pinf:=\liminf_{t\rightarrow \infty } p(t)$ and $\psup:=\limsup_{t\rightarrow \infty } p(t)$.
Equation \sysb{} can be written as $\dot y = p - \delta y$.
This is a linear system forced by the input $p=p(t)$.
Pick any $\varepsilon >0$.  Then there is some $T=T(\varepsilon )$ such that
$\pinf-\varepsilon <p(t)<\psup+\varepsilon $ for all $t\geq T$.
For such $t$, $\dot y(t)>0$ whenever $y(t)<(1/\delta )(\pinf-\varepsilon )$
and $\dot y(t)<0$ whenever $y(t)>(1/\delta )(\sup+\varepsilon )$.
It follows that $(1/\delta )(\pinf-\varepsilon )\leq y(t)\leq (1/\delta )(\pinf+\varepsilon )$ for all $t\geq T$.
Letting $\varepsilon \rightarrow 0$ we conclude that
\be{eq:limits1}
\pinf/\delta  \;\leq \; \liminf_{t\rightarrow \infty } y(t) \;\leq \; \limsup_{t\rightarrow \infty } y(t) \;\leq \; \psup/\delta 
\ee
and the inequalities \eqref{eq:ylims} follow when $a$$=$$b$$=$$c$$=$$1$.
To deal with general parameters, we recall that (\ref{rescaling:production}ab)
are obtained with $x = \frac{b}{a}\xstar$, $y = \frac{c}{b}\ystar$,
$t = \frac{1}{a}\tstar$, and $\delta ^* = \frac{\delta }{a}$.
Note that $\tstar\rightarrow \infty $ if and only if $t\rightarrow \infty $.
Thus \eqref{eq:limits1} holds 
for $p^* = u/\xstar = (b/a)p$, $\ystar$, and $\delta ^\star$
in place of $p$, $y$, and $\delta $.
Similarly, \eqref{eq:plims} holds for $p^* = u/\xstar$ and
\[
\muinf^* =  \liminf_{t\rightarrow \infty } v^*(\tstar)\,,\quad
\musup = \limsup_{t\rightarrow \infty } v^*(\tstar)\,,
\]
where
$v^*=\frac{du/d\tstar}{u} = (1/a) v$,
so $\muinf^*=(1/a)\muinf$ and $\musup^*=(1/a)\musup$.
Therefore,
\[
\liminf_{t\rightarrow \infty } y(t) \;=\;
\liminf_{\tstar\rightarrow \infty } \frac{c}{b}\ystar(\tstar) \;\geq \;
\frac{c}{b}\frac{\pinf^*}{\delta ^\star} \;=\; 
\frac{c}{b}\frac{\pinf^*}{\delta /a} \;=\;
\frac{ac}{b\delta } \,\pinf^* \;=\;
\frac{ac}{b\delta } \, \max\left\{0,1+\muinf^*\right\} \;=\;
\frac{c}{b\delta } \, \max\left\{a+\muinf\right\}
\,.
\]
A similar remark applies to $\limsup$, and the result follows.
\epr

\bc{cor:ylim}
If $v(t)\rightarrow \mu $ as $t\rightarrow \infty $ then 
$\displaystyle\lim_{t\rightarrow \infty } y(t) = \frac{c}{b\delta }\max\{0,a+\mu \}$.
\ec

Three particular cases are:
\bi
\item
When $u(t)$ has sub-exponential growth, meaning that $d\ln u/dt\leq 0$, then
$\displaystyle\limsup_{t\rightarrow \infty }y(t)\leq \frac{ac}{b\delta }$.
\item
In particular, if $u(t)=\alpha +\beta t$ is linear, then $\mu =0$ and thus
$\displaystyle\lim_{t\rightarrow \infty } y(t)=\frac{ac}{b\delta }$.
\item
If $u(t)=\beta e^{\mu t}$ is exponential, then
$\displaystyle\lim_{t\rightarrow \infty } y(t) = \frac{c}{b\delta }\max\{0,a+\mu \}$.
\ei

In conclusion, when $u$ is constant, or even with linear growth, the value of
the output $y(t)$ converges to a constant, which does not depend on the actual
constant value, or even the growth rate, of the input.
For constant inputs, this is called the ``perfect adaptation'' property.
If, instead, $u$ grows exponentially, then $y(t)$ converges to a steady state
value that is a linear function of the logarithmic growth rate.

\br{remark:sysdegradation.limits}
A possible alternative IFFL model is that in which $y$ follows this equation:
\be{sys:degrade}
\dot y \;=\; cu - \delta xy\,.
\ee 
instead of (\ref{sys:production.expo}b).
This model represents a different way of implementing the negative effect
of $x$ on $y$, through degradation instead of inhibition of production
A reduction to $a=b=c=1$ is again possible.
Now the substitutions
\be{rescaling:degradation}
x = \frac{b}{a}\xstar\,,\;
y = \frac{c}{a}\ystar\,,\;
t = \frac{1}{a}\tstar\,,\;
\delta ^* = \frac{b\delta }{a^2}\,,\;
\lambda ^* = \frac{\lambda }{a}\,,\;
\kappa ^* = \frac{c\kappa }{a^2}\,,
\ee
into (\ref{sys:production.expo}a-\ref{sys:degrade}-\ref{sys:production.expo}c)
transform the system into:
\bes{sys:degradation.expo.fixabc1}
\frac{d\xstar}{d\tstar} &=& -\xstar + u\\
\frac{d\ystar}{d\tstar} &=& u - \delta ^*\xstar\ystar\\
\frac{du}{d\tstar}  &=& (\lambda ^*-\kappa ^*\ystar) u
\ees
Consider a model that uses \eqref{sys:degrade} instead of
equation (\ref{sys:production.expo}b) and suppose that, for some $\gamma >0$,
$u(t)\geq \gamma >0$ for all $t\geq 0$
(for example, $u(t)=\beta e^{\mu t}$ or $u(t)=\alpha +\beta t$).
Then \eqref{eq:ylims} again holds, as does Corollary \ref{cor:ylim}.
This is because we one may rewrite
$\dot y = cu - c\delta y$ as $\dot y = x(cu/x - \delta y)$, and, provided that, for some $\xi >0$,
$x(t)>\xi >0$ for all $t$,
solutions have the same asymptotic behavior as for (\ref{sys:production.expo}b).
On the other hand, from the fact that $p(t)=u(t)/x(t)$ is bounded, we know
that, for some $\gamma '>0$, for all $t$, $x(t)\geq \gamma 'u(t)>\gamma '\gamma >0$.
\er

\section{IFFL's as feedback controllers}

As we remarked, in the case of exponential inputs
$u(t)=\beta e^{\mu t}$, 
$\displaystyle\lim_{t\rightarrow \infty } y(t) = \ybar = \frac{c}{b\delta }\max\{0,a+\mu \}$.
This holds both for \sysab{} and for the combination
\sysa-\eqref{sys:degrade}.
Now suppose that, in turn, $u(t)$ satisfies equation \sysc,
which means means that $v(t) = \lambda -\kappa y(t)$, and therefore
$\mu  = \lim_{t\rightarrow \infty }v(t) = \lambda -\kappa \ybar$.
This gives an implicit equation for the rate $\mu $:
\be{eq:muclosedloop}
\mu  = \lambda -\kappa \ybar = \lambda  - \frac{c\kappa }{b\delta }\max\{0,a+\mu \} \,.
\ee
We now solve this equation.

Suppose first that $a+\mu \geq 0$.  In that case, a solution has to satisfy
$\mu  = \lambda  - \frac{c\kappa }{b\delta }(a+\mu )$
and therefore there is a unique $\mu \geq  -a $ that solves the equation, namely:
\be{eq:solnm}
\mu  \;=\; \frac {\lambda b\delta -c\kappa a}{b\delta +c\kappa } \,.
\ee
Observe that $\mu \geq -a$ implies that 
$\lambda b\delta -c\kappa a \geq  (-a)(b\delta +c\kappa ) = -ab\delta  - ac\kappa $
and therefore $\lambda b\delta \geq -ab\delta $, or $\lambda \geq  -a$.
(And conversely, $\lambda \geq  -a$ implies $\lambda b\delta \geq -ab\delta $ and so 
$\lambda b\delta -c\kappa a \geq -ab\delta  - c\kappa a$ and hence $\mu \geq  -a$.)
So, if $\lambda  < -a$, there is no such solution.
Now we look for a solution with $a+\mu \leq 0$. 
Such an $\mu $ must satisfy $\mu =\lambda -0=\lambda $.
In summary, when $\lambda < -a$, the unique solution of \eqref{eq:muclosedloop}
is \eqref{eq:solnm}, and when $\lambda  \geq  -a$ it is $\mu =\lambda $.  

Note that when 
\be{eq:stability}
ca\kappa  \;>\; b\delta \lambda 
\ee
(which happens automatically when $\lambda <0$)
the formula \eqref{eq:solnm} gives that $m<0$, that is, $u(t)\rightarrow 0$ as $t\rightarrow +\infty $.
Conversely, if $ca\kappa <b\delta \lambda $, then $\mu >0$ and so $u(t)\rightarrow \infty $ as $t\rightarrow +\infty $.
Qualitatively, this makes sense: a large feedback gain $\kappa $, or a small growth
rate $\lambda $ in the absence of feedback, leads to the asymptotic vanishing
of the $u$ variable.

In addition, from the formula
$\ybar = \frac{c}{b\delta }\max\{0,a+\mu \}$ we conclude the following piecewise
linear formula for the dependence of the limit of the output on the parameter
$\lambda $ that gives the growth rate of $u$ when there is no feedback:
\be{eq:solnyclosedloop}
\ybar \;=\;
\twoif{0}{\mbox{ if } \; \lambda  < -a}%
      {\displaystyle\frac{c(a+\lambda )}{b\delta +c\kappa }} {\mbox{ if } \; \lambda  \geq  -a \,.}
\ee

These considerations provide helpful intuition about the closed-loop system,
but they do not prove that \eqref{eq:stability} is necessary and sufficient
for stability, nor do they show the validity of \eqref{eq:solnyclosedloop}
for the closed-loop system.  The reason that the argument is incomplete is
that there is no \emph{a priori} reason for $u(t)$ to have the exponential form
$u(t)=\beta e^{\mu t}$.  We next provide a rigorous argument.

\subsection{Analysis of the closed-loop system}

\bt{theorem:main}
Suppose that $(x(t),y(t),u(t)$ is a (positive) solution of \sysabc,
and define
\[
v(t)\,:=\;\dot u(t)/u(t)=\lambda -\kappa y(t)\,,
\]
\[
p(t)\,:=\;u(t)/x(t)\,,
\]
$\ybar$ by formula
\eqref{eq:solnyclosedloop}, which we repeat here:
\[
\ybar \;=\;
\twoif{0}{\mbox{ if } \; a+\lambda  < 0}%
      {\displaystyle\frac{c(a+\lambda )}{b\delta +c\kappa }} {\mbox{ if } \; a+\lambda  \geq  0}
\]
$\pbar:=(\delta /c)\ybar$. and
\[
\vbar \;=\;
\twoif{\lambda }{\mbox{if } \; a+\lambda  < 0}%
{\displaystyle \lambda  - \kappa \frac{c(a+\lambda )}{b\delta +c\kappa }} {\mbox{if } \; a+\lambda  \geq  0 \,.}
\]
Then:
\beqn
\lim_{t\rightarrow \infty }y(t) &=& \ybar\\
\lim_{t\rightarrow \infty }p(t) &=& \pbar\\
\lim_{t\rightarrow \infty }v(t) &=& \vbar\,.
\\
\lim_{t\rightarrow \infty }u(t)
&=&
\twoif {0} {\mbox{ if } \; ac\kappa  > b\delta \lambda }%
       {\infty }{\mbox{ if } \; ac\kappa  < b\delta \lambda  \,.}
\eeqn
\ets

\bpr
Substituting $v(t)=\lambda -\kappa y(t)$ into \eqref{eq:pv},
we have the surprising and very useful fact that there is a closed
system of just two differential equations for $p$ and $y$:
\bes{sys:py}
\dot p &=& p (a + \lambda -\kappa y -bp)\\
\dot y &=& cp - \delta y \,.
\ees
(This system could be viewed as a non-standard predator-prey of system, where
$y$ behaves as a predator and $p$ as a prey.)
In all of the real plane, there are two equilibria of this system, one at
$p=y=0$ and the other at $p=\frac{\delta (a+\lambda )}{b\delta +c\kappa }$,
$y=\frac{c(a+\lambda )}{b\delta +c\kappa }$. 
The second equilibrium point is in the interior of first quadrant if and only
if $a+\lambda >0$. 

We start by evaluating the Jacobian matrix of the linearized system.
This is:
\[
J = \mypmatrix{a+\lambda -\kappa y-2bp & -p\kappa \cr
c & -\delta }
\]
which, when evaluated at $p=y=0$, has determinant 
$-\delta (a+\lambda )$ and trace $a+\lambda -\delta $,
and when evaluated at $(\pbar,\ybar)$ has trace
\[
\frac{-b\delta (a+\lambda )}{c\kappa +b\delta }-\delta 
\]
and determinant $\delta (a+\lambda )$.
Thus, when $a+\lambda >0$, the trace is negative and the determinant is positive,
so the equilibrium $(\pbar,\ybar)$ is stable, and $(0,0)$ is a saddle because
the determinant of the Jacobian is negative at that point.
When instead $a+\lambda \leq 0$, the only equilibrium with non-negative coordinates is
$(0,0)$, and the determinant of the Jacobian is positive there, while the
trace is negative, so this equilibrium is stable.

We note that, in general, if have shown that there is a limit $v(t)\rightarrow \vbar$
as $t\rightarrow \infty $ then $u(t)\rightarrow 0$ as $t\rightarrow \infty $ if $\vbar<0$ and
$u(t)\rightarrow \infty $ as $t\rightarrow \infty $ if $\vbar>0$
Indeed, in the first case there is some $T\geq 0$ so
that for $t\geq T$, $v=\dot u/u < \vbar/2$, meaning that $d(e^{-\vbar t/2}u(t))/dt\leq 0$,
and hence $e^{-\vbar t/2}u(t)\leq e^{-\vbar T/2}u(T)$, so 
$u(t) \leq  e^{\vbar (t-T)/2}u(T)\rightarrow 0$ (since $\vbar<0$).
Similarly, in the second case we use that there is some $T\geq 0$ so
that for $t\geq T$, $v=\dot u/u > \vbar/2$, meaning that $d(e^{-\vbar t/2}u(t))/dt\geq 0$,
and hence $e^{-\vbar t/2}u(t)\geq e^{-\vbar T/2}u(T)$, so 
$u(t) \geq  e^{\vbar (t-T)/2}u(T)\rightarrow \infty $ (since $\vbar>0$).

Consider first the case $a+\lambda \leq 0$.
Then $\dot p = p(a + \lambda -\kappa y -bp)\leq p(-\kappa y-bp)<0$ for all $p>0$, and therefore
$p(t)\rightarrow \pbar=0$ as $t\rightarrow \infty $.
We may now view the linear system $\dot y=cp-\delta y$ as a one-dimensional system
with input $p(t)\rightarrow 0$, which implies that also $y(t)\rightarrow \ybar=0$.
In turn, this implies that $v=\lambda -\kappa y\rightarrow \vbar=\lambda <0$.
By the general fact proved earlier about limits for $u(t)$, we know that
$u(t)\rightarrow 0$ as $t\rightarrow \infty $.  This completes the proof when $a+\lambda \leq 0$.

So we assume from now on that $a+\lambda >0$.
We will show that, in this case, all solutions with $p(t)>0$ and $y(t)>0$
globally converge to the unique equilibrium $(\pbar,\ybar)$.
Once that this is proved, it will follow that $v(t)\rightarrow \vbar=\lambda -\kappa \ybar$.
Now, this value of $\vbar$, for $\ybar$ picked as in \eqref{eq:solnyclosedloop}
(case $a+\lambda \geq 0$), coincides with $\mu $ in \eqref{eq:solnm},
$\frac {\lambda b\delta -c\kappa a}{b\delta +c\kappa }$.  So $\vbar<0$ if $c\kappa a>\lambda b\delta $
and $\vbar>0$ if $\lambda b\delta >c\kappa a$, and this provides the limit statement
for $u(t)$, completing the proof.

\begin{figure}[ht]
\picc{0.4}{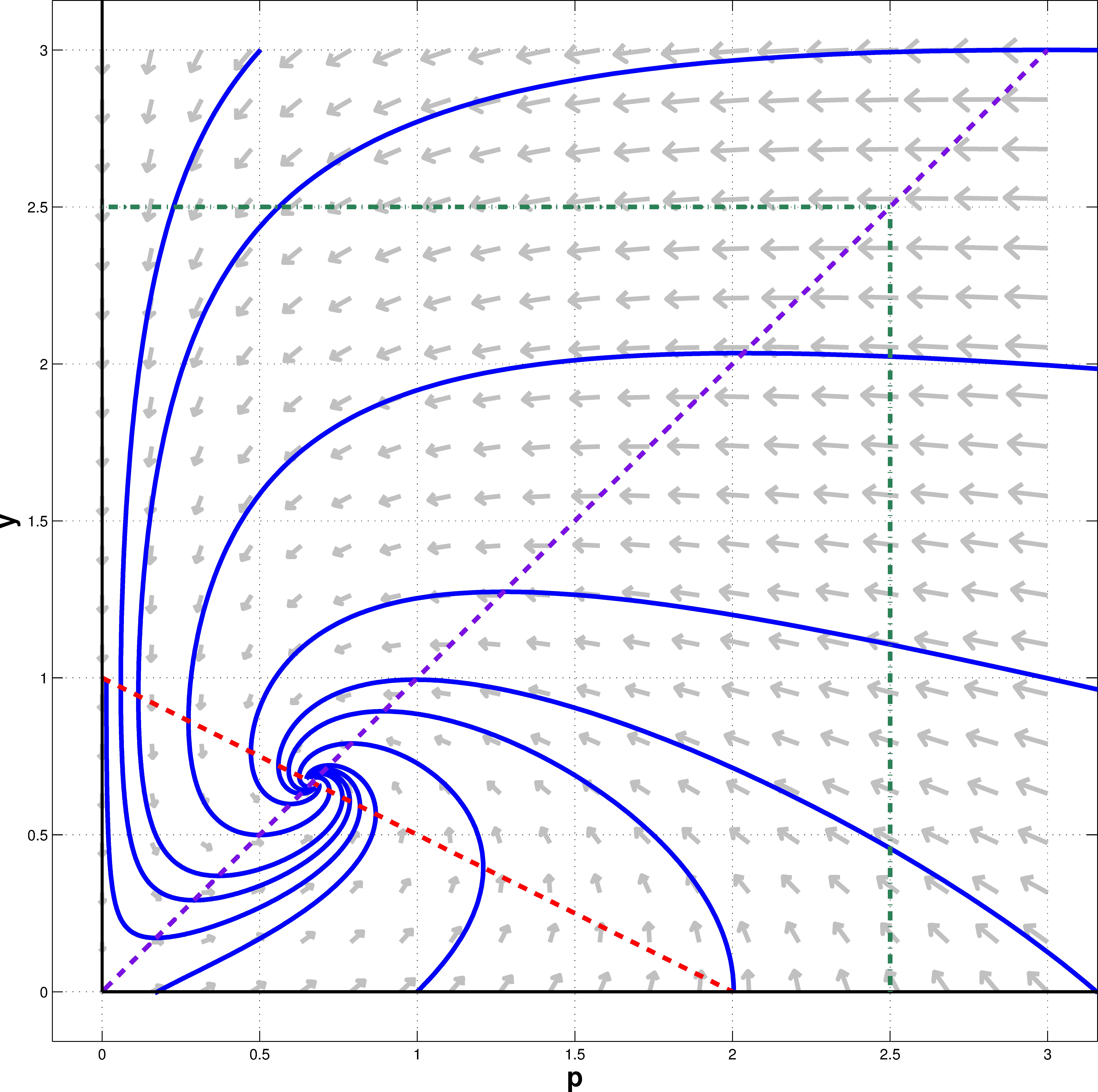}
\caption{Phase plane for \eqref{sys:py}, with several representative
trajectories plotted.  Nullclines are the $y$ axis,
corresponding to the stable manifold of $(0,0)$, and the lines given by
$y = (a+\lambda -bp)/\kappa $ (dashed red line) and $y=cp/\delta $ (dashed magenta line).  
In this plot, we picked $a=b=c=\lambda =\delta =1$ and $\kappa =2$, but the qualitative
picture is similar for all valid parameter values.
With these values, trajectories converge to the equilibrium
$(\pbar,\ybar)=(2/3,2/3)$.
Shown also is an invariant region $[0,P]\times [0,Y]$ with $P=Y=2.5$
(green dash-dotted lines and axes).
} 
\label{pplane1}
\end{figure}

We next show global convergence.  
A sketch of nullclines 
(see Fig.\ \ref{pplane1} for a numerical example)
makes convergence clear, and helps guide the proof.
Consider any $P\geq (a+\lambda )/b$ and any $Y\geq cP/\delta $ and
the rectangle $[0,P]\times [0,Y]$ (see Fig.\ \ref{pplane1}).
On the sides of this rectangle, the following properties hold:
\ben
\item
On the set $\{0\}\times (0,Y)$, $\dot p\geq 0$,
because $\dot p=0$.
\item
On the set $\{P\}\times (0,Y)$, $\dot p\leq 0$,
because $\dot p=p (a + \lambda - bP)\leq 0$, by the choice of $P$.
\item
On the set $(0,P)\times \{0\}$, $\dot y\geq 0$,
because $\dot y = cp>0$.
\item
On the set $(0,P)\times \{Y\}$, $\dot y\leq 0$.
because $\dot y = cp-\delta Y\leq cP-\delta Y\leq 0$ by the choice of $Y$.
\item
At the corner point $(0,0)$, $\dot p\geq 0$, $\dot y\geq 0$,
because $\dot p=\dot y=0$.
\item
At the corner point $(0,Y)$, $\dot p\geq 0$, $\dot y\leq 0$,
because $\dot p=0$, $\dot y=-\delta Y<0$.
\item
At the corner point $(P,0)$, $\dot p\leq 0$, $\dot y\geq 0$,
because $\dot p=p (a + \lambda - bP)\leq 0$, $\dot y=cP>0$.
\item
At the corner point $(P,Y)$, $\dot p\leq 0$, $\dot y\leq 0$,
because $\dot p=p (a + \lambda - bP - \kappa Y)<p (a + \lambda - bP)\leq 0$, $\dot y=cP-\kappa y\geq 0$. 
\een
These properties imply that the vector field points inside the set at every
boundary point and therefore it is forward-invariant, meaning that every
trajectory that starts in this set remains there for all positive times
\cite{nonsmooth-clarke-et-al-book}.
The rest of the proof of stability uses the  Poincar\'e-Bendixson Theorem
together with the Dulac-Bendixson criterion.
Note that, for any initial condition $\xi =(p(0),y(0))$ one can always pick a
large enough value of $P$ and $Y$ so that $(p(0),y(0))\in [0,P]\times [0,Y]$.
The invariance property guarantees that the omega limit set $\omega ^+(\xi )$ is a
nonempty compact connected set, and the Poincar\'e-Bendixson Theorem insures
that such a set is one of the following: (a) the equilibrium $(0,0)$, 
(b) a periodic orbit in the interior of the square, or (c) the equilibrium
$(\pbar,\ybar)$ \cite{HirschSmaleBook}.
Note that a homoclinic orbit around $(0,0)$ cannot exist, because the unstable
manifold of this equilibrium is the entire $y$ axis.  For the same reason,
if $\xi $ has positive coordinates, $\omega ^+(\xi )\not= (0,0)$.  Therefore, all that we
need to do is rule out periodic orbits.
Consider the function $\varphi(p,y)=1/p$.
The divergence of the vector field
\[
\mypmatrix{\frac{1}{p} (p(a + \lambda -\kappa y -bp))\cr
           \frac{1}{p}(cp - \delta y)}
\;=\;
\mypmatrix{a + \lambda -\kappa y -bp\cr
            c - \delta y/p}
\]
is $\frac{\partial a+\lambda -\kappa y-bp}{\partial p}
+
\frac{\partial c-\delta y }{\partial y}
=-b - \delta /p$, which has a constant sign (negative).
The Dulac-Bendixson criterion \cite{HirschSmaleBook} then guarantees that no
periodic orbits can exist, and the proof is complete.
\epr

\section{Adding positive feedback}

We now study a model in which there is an additional autocatalytic
positive feedback on $y$ variable.
We first consider the open loop system
\sysabp{},
and then discuss the full feedback system \sysabpc{}, which we repeat here for
convenience: 
\bes{eq:full_feedback_system}
\dot x &=& -ax + bu\\
\dot y &=& c\frac{u}{x} \,-\, \delta y \,+\, \frac{V y^n}{K^n + y^n}\\
\dot u &=& (\lambda -\kappa y)u
\ees

\subsection{Open-loop system with autocatalysis}

We first consider only the open-loop system 
\sysfeedbackab,
in which $u=u(t)$ is seen as an input function (stimulus) and $y$ as an output
(response). 

\begin{figure}[ht]
\picc{0.35}{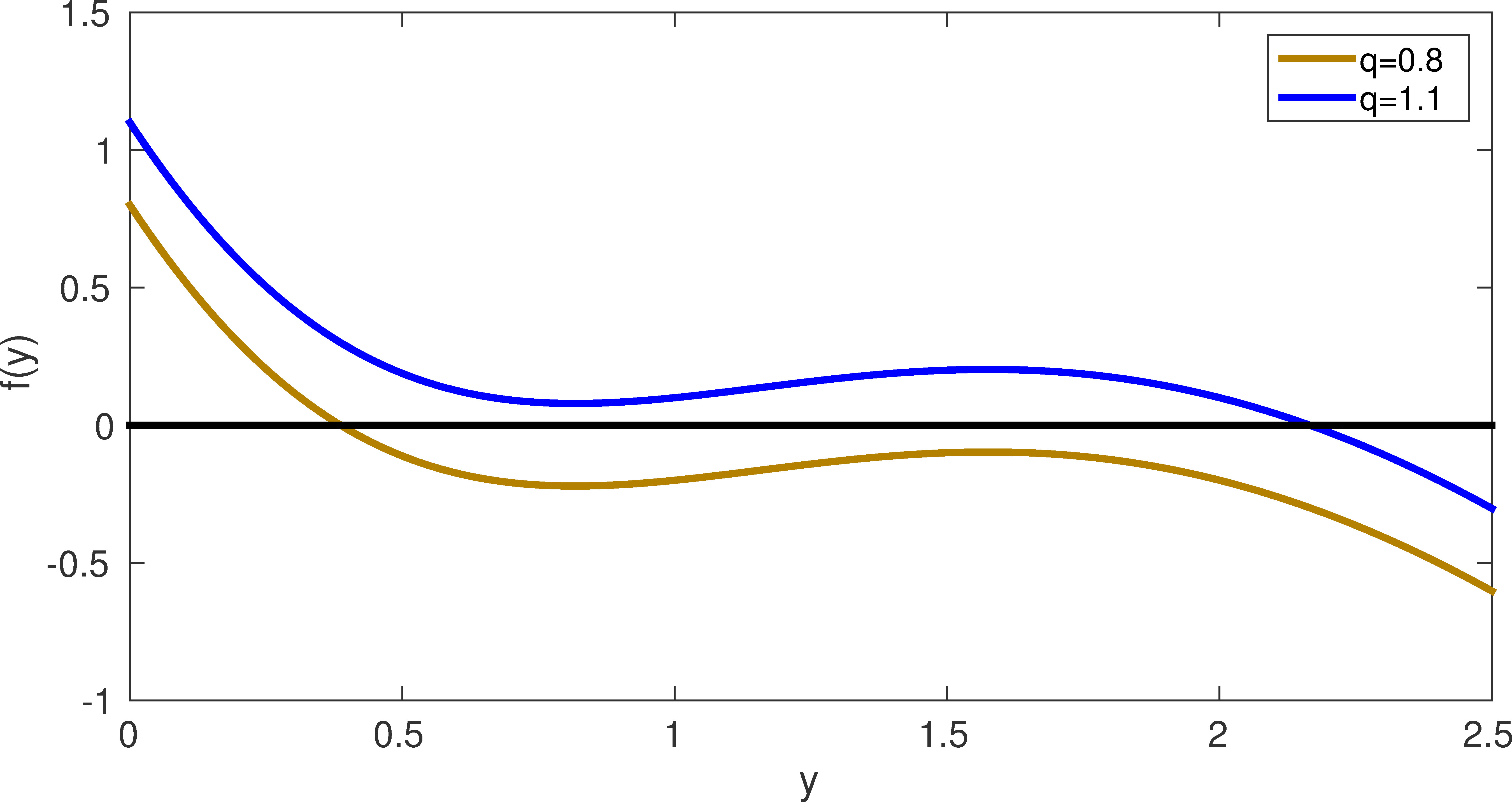}
\caption{Plots of $f(y)=q-\delta y+\frac{V y^n}{K^n + y^n}$, with
$a=b=c=1$, $\delta =3$, $n=2$, $V=10$, and $K=2$,
comparing $q=0.8$ (brown) and $q=1.1$ (blue).
The steady state changes from a low to a high value.
} 
\label{fig:bistable_two_qs}
\end{figure}

For appropriate parameters, and assuming that the Hill exponent (cooperativity
index) $n$ is greater than one, the system
\be{sys:ybistable}
\dot y \;=\; q \,-\, \delta y \,+\, \frac{V y^n}{K^n + y^n}\\
\ee
admits more than one steady state.
(In contrast, if there is no autocatalytic feedback, $V=0$, then there is a
unique steady state, $\ybar=q/\delta $.)
Let us fix all parameters except $q$, which we temporarily view as a bifurcation
parameter. 
Adjusting the value of $q$, one may obtain a low steady state, multiple
steady states, or a higher steady state.
As an illustration, pick $a=b=c=1$, $\delta =3$, $n=2$, $V=10$, and $K=2$.
Fig,~\ref{fig:bistable_two_qs} shows the right-hand side of
\eqref{sys:ybistable} plotted for $q=0.8$ and $q=1.1$.
For the latter value of $q$, there is larger steady state.
(Intermediate values typically give a system with two stable states and one
unstable state.)

Let us now write
$q(t) = c\frac{u(t)}{x(t)}$
in the system \sysfeedbackab.
Suppose that we consider an input $u$ which has a step increase at time $t=0$,
from $u(t)=u_-$ for $t<0$ to $u(t)=u_+$ for $t\geq 0$.
Suppose also that $x(0)=x_0=(b/a)u_-$, that is, that the system at time $t=0$ 
has an internal steady state preadapted to $u_-$.
Since $x(t)$ is a continuous function of time, we have that, for small times
$t>0$, $x(t)\approx x_0$ and $u(t)=u_0$, and thus
$q(t) \approx \alpha  u_+/u_-$, where $\alpha =ac/b$.
This means that the value of $q(t)$ for
$0\leq t \ll 1$ is proportional to the ``fold change'' in the input.
On the other hand, as $t\rightarrow \infty $, $x(t)\rightarrow b/a$, so $q(t)\rightarrow ac/b=\alpha $.
In the system with no autocatalytic effect ($V=0$), the differential equation
$\dot y = q-\delta y$ has a unique globally asymptotically stable equilibrium, and
therefore $y(t)\rightarrow q/\delta =\alpha /\delta $.
That is to say, there is 
\emph{complete adaptation}:
after a step increase in the input $u$, $y$ responds in a way that transiently
depends on the fold change, but it eventually returns to its adapted value.

\begin{figure}[ht]
\picc{0.33}{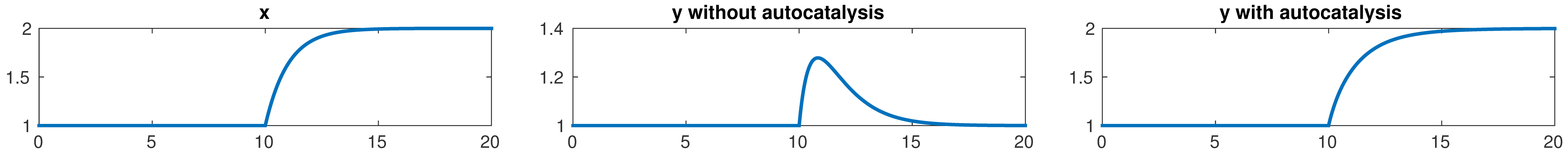}
\caption{Response to an input stepping from $u$$=$$1$ to $u$$=$$2$ (fold
  change of input is 2).  Comparing
  system with no positive feedback to system with positive feedback.
State $x(t)$ is the same in both systems, so only one panel is shown.
Parameters are $a=b=c=1$, $\delta =3$, $n=2$, $V=10$, and $K=2$ in system with
feedback, substituting $V=0$ in system without feedback.
} 
\label{fig:bistable_vs_linear_response}
\end{figure}

On the other hand, if there is an autocatalytic feedback term ($V\not= 0$), the
initial input $q(t)$ to the $y$-subsystem may trigger an irreversible
transition to a different state $y$ than the adapted value.  
Since the initial value of $q(t)$ depends on the fold change of the input,
this implies that for different ranges of fold-change magnitudes, $y$ might
switch to different states, and remain there even after the excitation goes
away. 
As an example, using the same parameters $a=b=c=1$, $\delta =3$, $n=2$, $V=10$, and
$K=2$ as earlier, Fig,~\ref{fig:bistable_vs_linear_response} shows how
a step change in the input can result in an irreversible locking to a higher
activation state, for the system with feedback, compared with the system without
feedback, which does not switch but has only a transient change in activity.

\subsection{Closed-loop system with autocatalysis}
\begin{figure}[ht]
\picc{0.5}{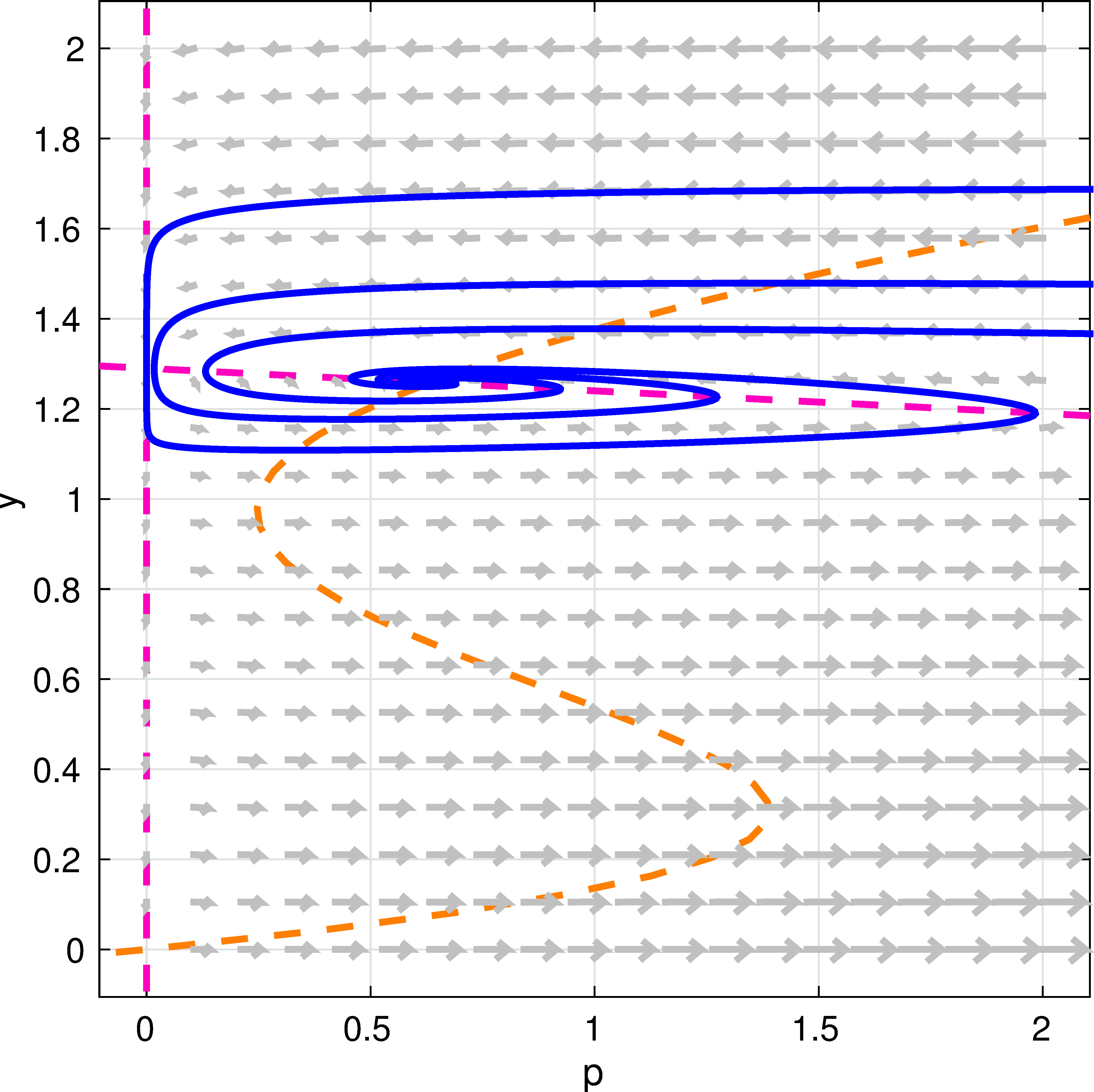}
\caption{Phase-plane for system \eqref{sys:pyfeedback}, with
$a=0.8$,
$b=1$,
$c=0.1$,
$\delta =1$,
$n=2$,
$V=1.95$,
$K=1$,
$\kappa =20$
$\lambda =25$.
The $y$-nullcline is
$cp - \delta y + \frac{V y^n}{K^n + y^n}=0$ (dot-dashed orange).
The $p$-nullcline has two components: $p=0$ (the $y$-axis) and the line
$y = (a+\lambda -bp)/\kappa $ (dashed red).
Three representative trajectories are shown (solid blue).
Notice the vertical-looking motion of one trajectory near the $y$-axis: along
such solutions, $p(t)=u(t)/x(t)$ stays $\approx0$ for a time interval, after
which this ratio converges to $\pbar$.
Gray arrows indicate directions of movement in phase plane.
The equilibrium point $(\pbar,\ybar)$ is such that
$p<0.8$ and thus,
since $\dot u=(\lambda -\kappa y)u$, $u(t)$ behaves like $\alpha e^{\mu t}$ for large $t$,
where $\mu =\lambda -\kappa \ybar=bp-a =p-0.8$,
we have $\mu <0$ (elimination).
}
\label{fig:pyfeedback_phaseplane}
\end{figure}

We now turn to the full feedback system
\sysfeedbackabc.
Just as in the case in which there was no autocatalytic terms, 
we may again reduce to a two-dimensional system written in terms of
$p=u/x$ and $y$.  The system is now:
\bes{sys:pyfeedback}
\dot p &=& p (a + \lambda -\kappa y -bp)\\
\dot y &=& cp - \delta y + \frac{V y^n}{K^n + y^n}\,.
\ees
For appropriate parameter regimes, there is a unique positive steady state
$(\pbar,\ybar)$.
Specifically, for $n>1$ the derivative of
$\frac{V y^n}{K^n + y^n}$
attains its maximum at
$y = (\frac{n-1}{n+1})^{1/n} K = K/\sqrt3$
when $n=2$, and the derivative is
$\frac{3V\sqrt3}{8K}$ there.
Thus, the function 
\[
g(y) = a + \lambda -\kappa y -(b/c) \left(\delta y - \frac{V y^n}{K^n + y^n}\right) \,,
\]
whose roots determine the nonzero equilibrium values of $y$,
has derivative $\leq  - \kappa  - b\delta /c + \frac{3V\sqrt3}{8K}$.
Thus, when
\[
\frac{3V\sqrt3}{8K}<\kappa  + b\delta /c
\]
the function $g$ is strictly
decreasing and therefore (in the nontrivial case $a+\lambda >0$),
since $g(0)>0$ and $g(y)\rightarrow -\infty $ as
$y\rightarrow \infty $, there is a unique zero $\ybar$. 
See for example the
phase plane drawn in Fig.~\ref{fig:pyfeedback_phaseplane}.

A remarkable feature emerges for this system.
When does $u(t)\rightarrow 0$ as $t\rightarrow \infty $, corresponding to elimination of a pathogen or
tumor, in the motivating context of immunology?  
When does $u(t)\rightarrow \infty $ as $t\rightarrow \infty $, corresponding to proliferation?  
Note that, if
$(p(t),y(t))\rightarrow (\pbar,\ybar)$ as $t\rightarrow \infty $, then,
since $\dot u=(\lambda -\kappa y)u$, $u(t)$ behaves like $\alpha e^{\mu t}$ for large $t$.
On the other hand, at steady state $a + \lambda -\kappa \ybar -b\pbar=0$, which means
that 
$\mu -\lambda -\kappa \ybar =b\pbar-a$. Therefore:
\beqn
\pbar \,<\, \frac{a}{b} &\Rightarrow & u(t) \rightarrow  0 \;\mbox{as}\, t\rightarrow \infty \\
\pbar \,>\, \frac{a}{b} &\Rightarrow & u(t) \rightarrow  \infty  \;\mbox{as}\, t\rightarrow \infty \,.
\eeqn

\begin{figure}[ht]
\picc{0.5}{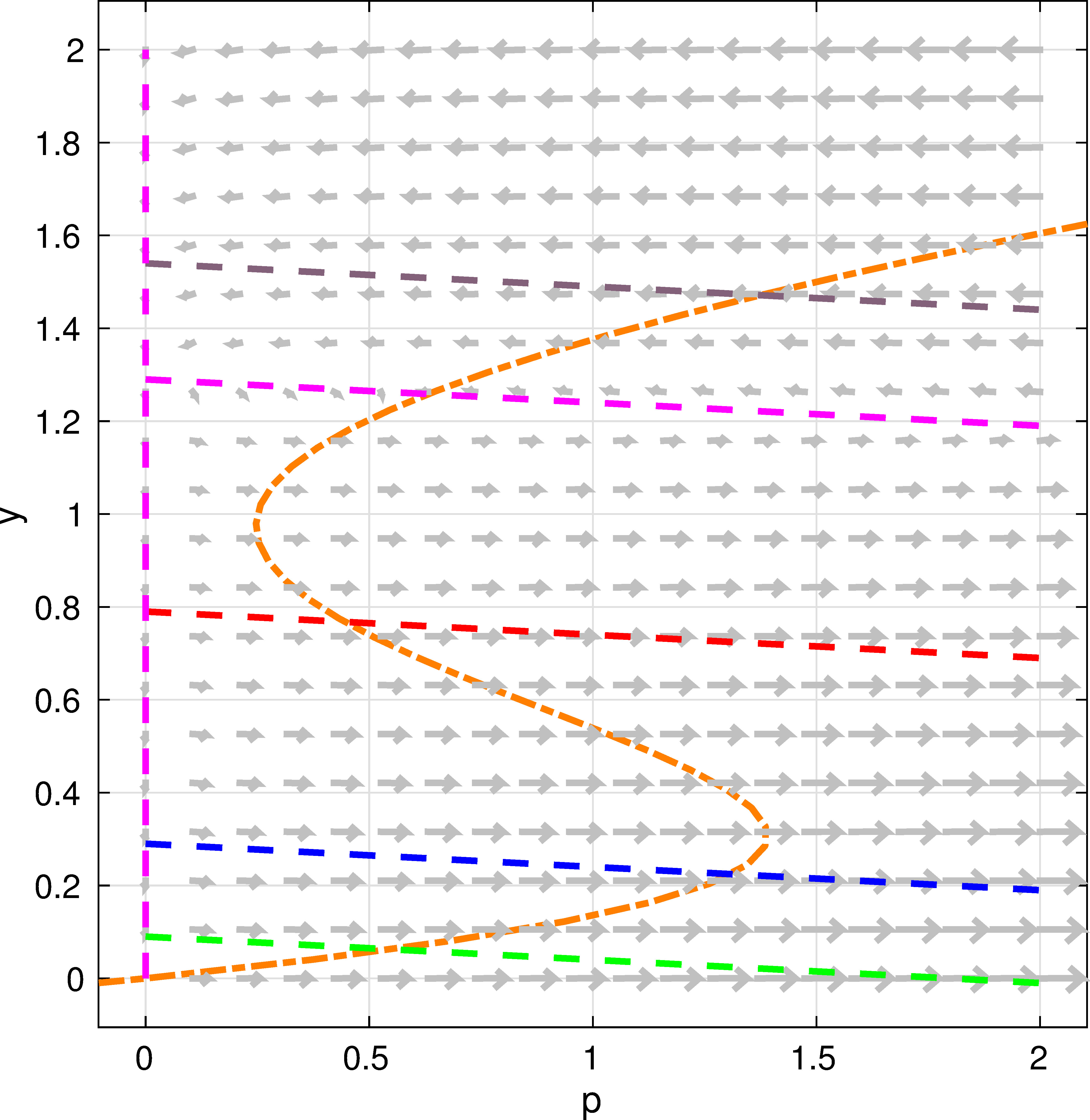}
\caption{Phase plane for system \eqref{sys:pyfeedback}, with
$a=0.8$,
$b=1$,
$c=0.1$,
$\delta =1$,
$n=2$,
$V=1.95$,
$K=1$,
$\kappa =20$,
same parameters as in Fig.~\ref{fig:pyfeedback_phaseplane}, but now with
several values of $\lambda $.
The $y$-nullcline is
$cp - \delta y + \frac{V y^n}{K^n + y^n}=0$ (dot-dashed orange).
The $p$-nullcline has two components: one is $p=0$ (the $y$-axis, dashed
magenta) and the second component is the line $y = (a+\lambda -bp)/\kappa $
which depends on the value of $\lambda $, and is
shown for
$\lambda =1$ (green),
$\lambda =5$ (blue),
$\lambda =15$ (red),
$\lambda =25$ (magenta),
and $\lambda =30$ (black).
Gray arrows indicate directions of movement in phase plane for $\lambda =25$.
Observe that the equilibrium point $(\pbar,\ybar)$ is such that
$p<0.8$ for $\lambda =1$,
$p>0.8$ for $\lambda =5$,
$p<0.8$ for $\lambda =15$ and $\lambda =25$,
and $p>0.8$ for $\lambda =30$.
Since $\dot u=(\lambda -\kappa y)u$, $u(t)$ behaves like $\alpha e^{\mu t}$ for large $t$,
where $\mu =\lambda -\kappa \ybar=bp-a =p-0.8$,
these growth rates $\lambda $ corresponds respectively to
$\mu <0$ (elimination), 
$\mu <0$ (proliferation), again
$\mu <0$ (elimination, two nullclines values shown), and yet again
$\mu >0$ (proliferation).
}
\label{fig:pyfeedback_several_nullclines}
\end{figure}

\begin{figure}[ht]
\picc{0.3}{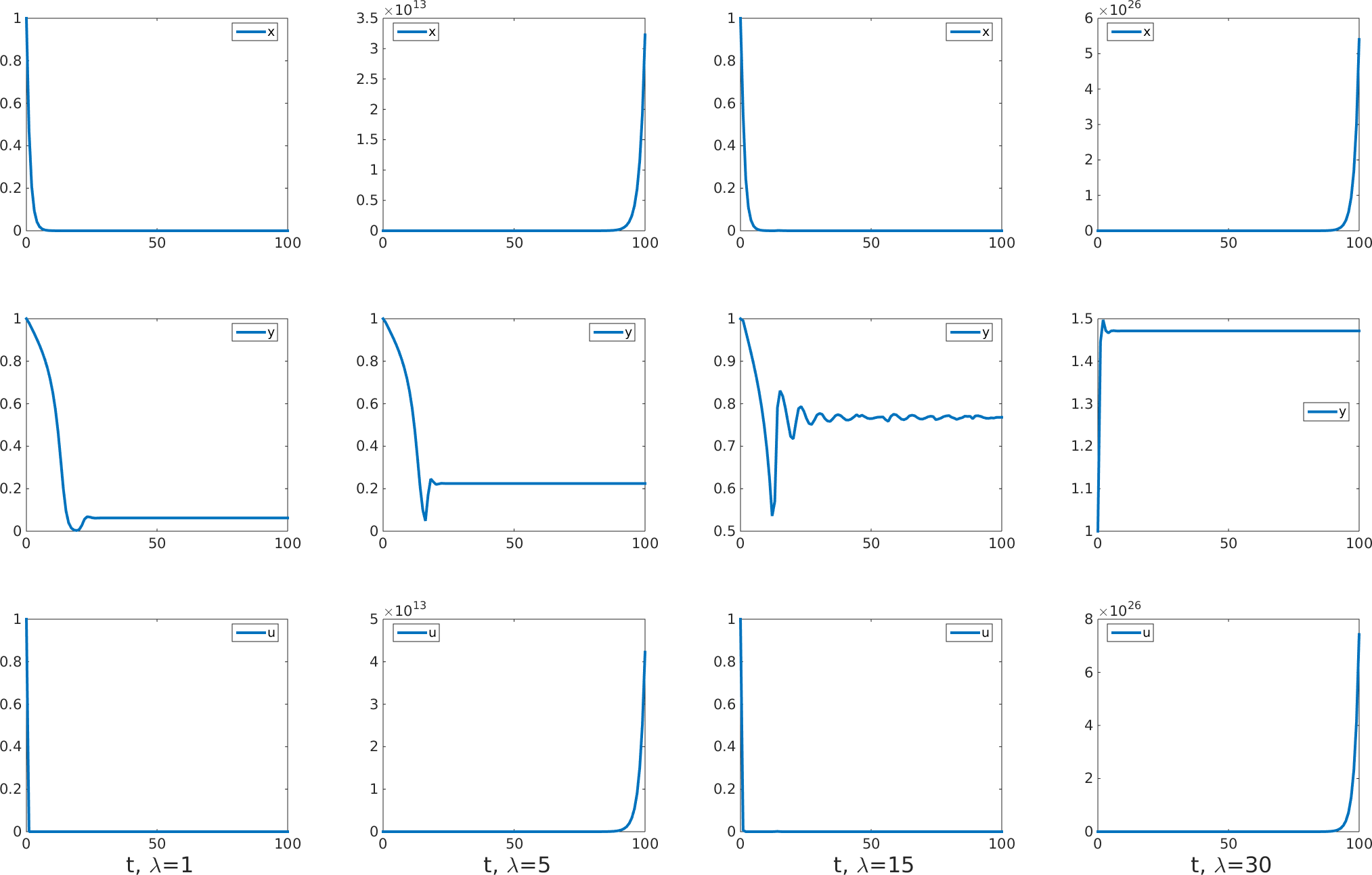}
\caption{Simulations of system \eqref{sys:pyfeedback}, with 
$a=0.8$,
$b=1$,
$c=0.1$,
$\delta =1$,
$n=2$,
$V=1.95$,
$K=1$,
$\kappa =20$,
same parameters as in Fig.~\ref{fig:pyfeedback_phaseplane}, but now with
several values of $\lambda $.
Initial states $x(0)=y(0)=u(0)=1$.
Left to right: $\lambda =1, 5, 15, 30$.
Asymptotic behavior of $u(t)$ shown in bottom panel is as expected from
Fig.~\ref{fig:pyfeedback_several_nullclines}.
As the growth rate $\lambda $ increases, we obtain 
elimination (for $\lambda =1$), 
proliferation ($\lambda =5$),
elimination ($\lambda =15$; plot for $\lambda =25$ not shown but similar), and again 
proliferation ($\lambda =30$)
} 
\label{fig:pyfeedback_samples}
\end{figure}

\begin{figure}[ht]
\picc{0.8}{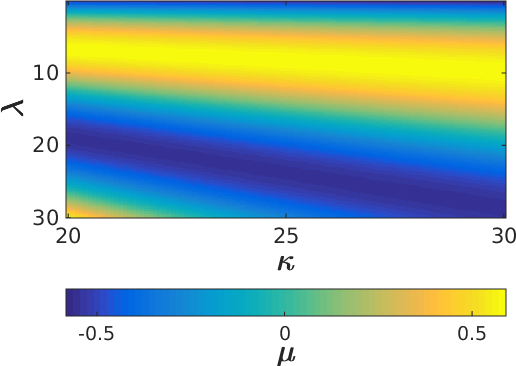}
\caption{Values of the effective rate $\mu =\lambda -\kappa \ybar$ at steady state.
  Colors scale shown at bottom.
  Values are obtained by algebraically solving for a steady state
  $(\pbar,\ybar)$ in system   \eqref{sys:pyfeedback} and then computing
  $\mu =\lambda -\kappa \ybar$.  Since $\dot u=\mu  u$, it follows that $u(t)$ behaves like
  $\alpha e^{\mu t}$ for large $t$. 
  Therefore $\mu <0$ corresponds to $u(t)\rightarrow 0$ as   $t\rightarrow \infty $ and 
  $\mu >0$ to $u(t)\rightarrow \infty $ as $t\rightarrow \infty $.
  Parameters are:
$V=1.95$,
$K=1$,
$\delta =1$,
$a=0.8$,
$b=1$,
$c=0.1$,
as in Fig.~\ref{fig:pyfeedback_phaseplane}, but now showing the effect of
varying both $\kappa $ and $\lambda $.
For $\kappa =20$, 
as the growth rate $\lambda $ increases from $0$ to $30$ (going down a column), we
obtain elimination, proliferation, again elimination, and finally again 
proliferation
} 
\label{fig:pyfeedback_heatmap}
\end{figure}

Note that
$(\pbar,\ybar)$ is a positive equilibrium if and only if
$b\pbar = a + \lambda -\kappa \ybar$ and $c\pbar = -f(\ybar)$.
To find equilibria, we can first solve
$a + \lambda -\kappa \ybar=-(b/c)f(\ybar)$ for $\ybar$, and then obtain
$\pbar$ simply as $(1/b) (a + \lambda -\kappa \ybar)$.
Note that 
$\mu =\lambda -\kappa \ybar<0$ is equivalent to $\pbar>a/b$, or $-(1/c)f(\ybar)>a/b$, and
$\mu =\lambda -\kappa \ybar<0$ is equivalent to $\pbar<a/b$, or $-(1/c)f(\ybar)<a/b$.
Therefore, leaving all other parameters constant, $\mu $ switches sign whenever
$f(\ybar)=-ac/b$.
The formula
\[
\lambda  \;=\; \kappa \ybar - (b/c)f(\ybar) - a
\]
gives those values of $\lambda $ where there is change from
$\mu <0$ (which means $u(t)\rightarrow 0$ as $t\rightarrow \infty $)
to $\mu >0$ ($u(t)\rightarrow \infty $ as $t\rightarrow \infty $),
or viceversa.
\emph{As $\lambda $ increases, we may expect several such switches}, as may be
seen graphically as one draws parallel nullclines corresponding to
different values of $\lambda $.  For the  
example in Fig.~\ref{fig:pyfeedback_phaseplane}, several of these are shown
in Fig.~\ref{fig:pyfeedback_several_nullclines}.
Simulations confirm these phase planes,
see Fig.~\ref{fig:pyfeedback_samples}.
The heatmap in Fig.~\ref{fig:pyfeedback_heatmap} shows graphically how various
combinations of $\lambda $ and $\kappa $ lead to growth or elimination, for these
parameters.

The ranges of growth rates $\lambda $ for which each of the intermediate
proliferation and elimination regimes can hold could be quite large. 
To illustrate how large these ranges could potentially be, consider the following
parameters: 
$b=1$,
$c=0.1$,
$\kappa =20$
$K=1$,
$V=2$,
$\delta =1$,
and $a=0.1$.
There is then a more than three order of magnitude range of $\lambda $'s (from
$\lambda \approx0.004$ to $\lambda \approx17$) for which $u(t)\rightarrow \infty $, but a larger $\lambda $
results in elimination of $u$ (up to $\lambda \approx27$, after which again
$u(t)\rightarrow \infty $).
As another example, letting $a=1.2$, we find that there is an over four-fold
possible change in $\lambda $ (from $\lambda \approx1.4$ to $\lambda \approx6$) that results in
$u(t)\rightarrow \infty $, followed by another over four-fold possible change in $\lambda $ 
(from $\lambda \approx6$ to $\lambda \approx28$) 
that results in $u(t)\rightarrow 0$ (after which again $u(t)\rightarrow \infty $).

\clearpage
\newpage
\bibliographystyle{plain}

\begin{thebibliography}{1}

\bibitem{nonsmooth-clarke-et-al-book}
F.H. Clarke, Y.S. Ledyaev, R.S. Stern, and P.R. Wolenski.
\newblock {\em Nonsmooth Analysis and Control Theory (Graduate Texts in
  Mathematics}.
\newblock Springer-Verlag, New York, 1998.

\bibitem{HirschSmaleBook}
M.~W. Hirsch and S.~Smale.
\newblock {\em Differential Equations, Dynamical Systems and Linear Algebra}.
\newblock Academic Press, 1974.

\bibitem{biorxiv_change_detection_immune_2015}
E.D. Sontag.
\newblock Incoherent feedforward motifs as immune change detectors.
\newblock Technical report, bioRxiv http://dx.doi.org/10.1101/035600, December
  2015.

\end{thebibliography}

\edo